\documentclass[twocolumn,preprintnumbers,amsmath,amssymb]{revtex4}

\usepackage{graphicx}
\usepackage{dcolumn}
\usepackage{bm}
\usepackage[latin1]{inputenc} \newcommand{\comment}[1]{}

\begin{document}
\setlength{\unitlength}{0.7\textwidth} \preprint{}

\title{On the unsteady behavior of turbulence models}

\author{R. Rubinstein{$^1$} and W.J.T. Bos{$^2$}}

\affiliation{$^1$Newport News, Virginia, USA\\
$^2$Universit\'e de Lyon - LMFA - CNRS UMR 5509 - Ecole Centrale de Lyon - UCBL - INSA Lyon, Ecully, France}

\begin{abstract}
Periodically forced turbulence is used as a test-case to evaluate the
predictions of two-equation and multiple-scale turbulence models in
unsteady flows. The limitations of the two-equation model are shown to
originate in the basic assumption of spectral equilibrium. A multiple-scale
model based on a picture of stepwise energy cascade overcomes some of these
limitations, but the absence of nonlocal interactions proves to lead to
poor predictions of the time variation of the dissipation rate. A new
multiple-scale model that includes nonlocal interactions is proposed
and shown to reproduce the main features of the frequency response correctly.
\end{abstract}

\maketitle

A basic premise of 
one point closures such as the $k-\epsilon$ model is
the hypothesis of ``spectral equilibrium," which justifies two distinct
roles of the dissipation rate
$\epsilon$: on the one hand, it appears in the energy
balance, defined as a correlation of velocity gradients, 
hence a small-scale quantity; on the other hand, it is used
phenomenologically to describe large-scale transport properties. 
The most basic formulation of the latter is Kolmogorov's hypothesis
$\epsilon \propto k^{3/2}/L$ where $k$ is the turbulent kinetic energy
and where $L$ is a length scale
characteristic of the largest scales of motion; 
equivalent formulations include $\nu_t \propto k^2/\epsilon$, the
formula for turbulent viscosity, or $\tau \propto k/\epsilon$, the
formula for the turbulent time-scale.
The Kolmogorov theory, or more general assumptions of self-similarity
of all scales of motion, 
justify all of these proportionalities
\cite{Oberlack1999}, although the constants
of proportionality need not coincide in all self-similar flows
\cite{Bos2007-2}.

But turbulence models are not needed to describe self-similar
flows, which merely serve as calibration cases; 
models are needed to describe departure from self-similarity,
when spectral equilibrium becomes a strong
constraint on turbulence evolution. In a study of a flow in which
turbulence evolves from a steady state to a self-similar time-dependent
state \cite{Bob04}, the implications of the
departure from spectral equilibrium were investigated: this departure 
was connected to
transient failure of the Tennekes-Lumley balance \cite{Tennekes} and to
the consequent relevance of small scale dynamics for the large scales.
The breakdown of spectral equilibrium has also been observed in
engineering flows including
turbulent diffusers and wakes.\cite{Touil3}

The limitations of the spectral equilibrium hypothesis are very
well known in the modeling literature 
and have led to proposals for {\em multiple-scale models} \cite{Schiestel1987}
that more realistically address the complexity of the nonlinear
interactions in turbulent flows.
This letter reports on some investigations of
multiple-scale models applied to an especially simple and attractive
test case for transient turbulence: {\em periodically forced turbulence}
\cite{Lohse2000,Cadot2003,Kuczaj2008}. 
This problem arises when isotropic incompressible
turbulence, maintained in a steady state by a large-scale isotropic forcing 
with total amplitude $\bar p$, 
is subjected to a small time-dependent periodic perturbation with amplitude $\tilde p$:
$p(t)=\bar p+\tilde p~ \cos(\omega t)$, 
such that the ratio $\tilde p/\bar p \ll 1$ 
and such that the forcing length scale does not depend on time. 
The phase-averaged kinetic energy
$k$
can then be 
decomposed into a mean $\bar k$ and a periodic part 
$\tilde k(\omega)\cos(\omega t+\phi_k(\omega))$, with $\phi_k$ the phase-shift 
between the forcing and the kinetic energy. 
Similarly, the viscous dissipation rate can be written as 
$\epsilon=\bar \epsilon+\tilde \epsilon(\omega)\cos(\omega t+\phi_\epsilon(\omega))$;
$\bar k$ and $\bar\epsilon$ are related to the time-independent
forcing length scale $\bar{L}$ by $\bar{L}\propto {\bar k}^{3/2}/\bar{\epsilon}$. 
The periodic parts of $k$ and $\epsilon$ are sinusoidal, like the forcing, 
because $\tilde p/\bar p\ll 1$. 
The functions $\tilde k(\omega)$, $\tilde \epsilon(\omega)$, 
$\phi_k(\omega)$, and $\phi_{\epsilon}(\omega)$, which
characterize the linearized response of 
steady-state turbulence to periodic perturbation of the forcing, 
can be called the {\em linear response functions}.

We will use periodically forced turbulence as a test case to evaluate 
the ability of multiple-scale models to predict the dynamics of time-dependent turbulence. 
It will first be shown that the unsteady predictions of multiple-scale models 
are significantly better than the predictions of
a two-equation model. However, an elementary multiple-scale model based on the
heuristic picture of stepwise energy cascade is found to have limitations in
predicting the unsteady dissipation rate. A multiple-scale model that includes
the possibility of nonlocal interactions is proposed; it is shown that this model
can capture some fine features of the unsteady energy dissipation.

The linear response functions were determined in recent work \cite{Bos2007-3}
using the EDQNM closure theory, which
was shown
to compare very well to available low Reynolds number
experimental, and DNS data.
Comparison with high Reynolds number data for linear response functions 
would be desirable, but such data is not yet available.
Briefly summarizing the major conclusions,
the two-equation model is satisfactory
both in the {\em static} limit $\omega\rightarrow 0$, in which
the phase shifts $\phi_k,\phi_{\epsilon}$ vanish, and 
in the {\em frozen} limit $\omega\rightarrow\infty$, in which 
$\tilde{k}\sim\omega^{-1}$ and $\phi_k\sim\pi/2$. 
However, this agreement is trivial;
only the results at intermediate frequencies 
provide a real test of the model. 
The two-equation model reproduces
the  function $\tilde{k}(\omega)$ reasonably well, 
but the transition from the static limit $\phi_k\approx 0$
to the frozen limit  $\phi_k\approx\pi/2$ occurs over a frequency
range that is much too wide.
The amplitude $\tilde{\epsilon}(\omega)$ is not satisfactory:
a range in which $\tilde{\epsilon}\sim\omega^{-3}$ at high Reynolds numbers is 
absent. We note that since observations of the
modulated dissipation rate are very difficult, and relevant high
Reynolds number data is not yet available,
the EDQNM results for this quantity remain
theoretical predictions; they are nevertheless
supported by arguments \cite{Bos2007-3} based on the well-established
role of distant interactions in turbulence.
Finally, the two-equation model also makes the incorrect prediction that 
$\phi_k=\phi_{\epsilon}$ regardless of $\omega$.
\begin{figure}
\setlength{\unitlength}{1\textwidth}
\includegraphics[width=0.25\unitlength]{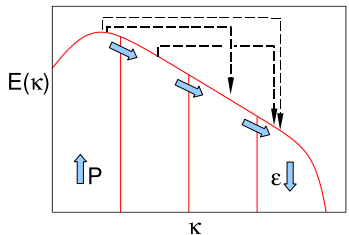}
\caption{Schematic view of a discretized energy cascade. The solid arrows indicate the energy fluxes between neighboring wavenumber shells. The dashed lines indicate the nonlocal fluxes between spectrally remote shells. Note that a similar picture can be found in a paper by Lumley \cite{Lumley92}.
\label{Cartoon}}
\end{figure}

We investigate whether these predictions can be improved using multiple-scale modeling
following the ideas of Schiestel \cite{Schiestel1987}.
Multiple-scale models can be considered numerical 
methods for spectral closures, with significant modifications designed
to permit reasonable accuracy at a very low order of discretization.
Thus, whereas a numerical implementation of a spectral closure would
solve for the energy spectrum at perhaps hundreds of 
discrete wavenumbers $\kappa_i$, in Schiestel's formulation,
the energy spectrum is divided 
into a relatively small number of wavenumber shells $\kappa_{i-1}\leq\kappa\leq\kappa_i$;
for each shell, equations are 
written for two scalar descriptors: the fluctuation energy contained
in the shell and the net energy flux into it.
To enhance the accuracy possible
with a relatively small number of
shells, Schiestel allowed the partition wavenumbers $\kappa_i$ to be functions of time.
A schematic picture of the resulting discretized
energy cascade is given in Figure \ref{Cartoon}, following Lumley \cite{Lumley92}.

The starting point for the analytical formulation
is the Lin equation governing the energy spectrum $E(\kappa,t)$,
\begin{equation}\label{Lin}
\left(\frac{\partial}{\partial t}+\nu \kappa^2 \right)E(\kappa,t)=P(\kappa,t)-
\frac{\partial F(\kappa,t)}{\partial \kappa}.
\end{equation}
In this equation $\nu$ is the viscosity, $P(\kappa,t)$ the forcing term, 
and $F(\kappa,t)$ is the energy flux across wavenumber $\kappa$. 
Schiestel applied the Kovaznay model \cite{Monin} 
\begin{equation}\label{eqKov}
F(\kappa,t)=C\kappa^{5/2} E(\kappa,t)^{3/2},
\end{equation}
with $C$ a parameter which determines the Kolmogorov constant. 
This model represents a stepwise cascade of energy in spectral space from small 
to large $\kappa$ and reproduces a $\kappa^{-5/3}$ inertial range. 
Figure \ref{Cartoon} would correspond to a stepwise cascade if the dashed 
lines were absent. 

To obtain a multiple-scale model, 
write the equation for the time derivative of 
the spectral energy flux $F(\kappa,t)$,
\begin{equation}\label{dotF}
\dot F(\kappa,t)
=\frac{3}{2} \frac{F(\kappa,t)}{E(\kappa,t)} \dot E(\kappa,t).
\end{equation}
From the viewpoint of Schiestel's analysis, we have assumed that
the partition wavenumbers are constant in time: this assumption seems
appropriate for this problem, in which the forcing wavenumber is fixed.
The production scales are assumed to be confined to low $\kappa$, and the 
dissipation to high $\kappa$ (a more general description including finite-Reynolds 
number effects or broad-band forcing will not be attempted here). 
For $\kappa$ in the inertial range, (\ref{Lin}) becomes
\begin{equation}\label{dotE}
\dot E(\kappa,t)=-\frac{\partial F(\kappa,t)}{\partial \kappa}.
\end{equation}
We combine this with (\ref{eqKov}) and (\ref{dotF}) to obtain
\begin{equation}
\dot F(\kappa,t)=-\frac{3}{2}\frac{F(\kappa,t)}{E(\kappa,t)}
\frac{\partial F(\kappa,t)}{\partial \kappa}.
\label{dotF1}
\end{equation}
A discrete model is obtained by splitting the spectral domain into $n$ 
shells as illustrated in Figure \ref{Cartoon}. 
The energy in shell $i$ is $e_i\approx E(\kappa_i)\Delta \kappa_i$. 
The spectral flux $f_i\approx F(\kappa_i)$ and the time derivative is 
$\partial F(\kappa_i,t)/\partial k\approx (f_i-f_{i-1})/\Delta \kappa_i$, 
so that we obtain 
\begin{equation}\label{Schiestel}
\dot f_i=-\frac{3}{2}\frac{f_i}{e_i}(f_i-f_{i-1}), \
\ 1\leq i\leq n.
\end{equation}
Integrating (\ref{dotE}) over each shell gives the partial energy balance
equations
\begin{equation}\label{Schiestel-energy}
\dot e_i=-(f_i-f_{i-1}), \
\ 1\leq i\leq n.
\end{equation}
in which all $f_i$ and $e_i$ are functions only of time. 
The oscillating production term $\tilde{p}$, assumed to act at the small wavenumbers, is identified 
with $f_0$, the flux entering shell $1$. 
Furthermore, the high Reynolds numbers case is considered in which we assume 
that the viscous dissipation takes place at the last wavenumber shell: 
$\epsilon=f_n$.  This assumption makes it unnecessary to introduce partial
dissipation rates for each shell and corresponding equations of motion.
A special feature of the Kovaznay model is that
the partition wavenumbers do not appear in the model.

By choosing $n=1$, one obtains a two-equation model; the equation for $f_1$ becomes the dissipation rate equation $\dot{\epsilon} = (3/2)(\epsilon/k)(P-\epsilon)$. Note that the two model constants, generally called $C_{\epsilon 1}$ and $C_{\epsilon
  2}$ in the literature, are equal, which allows the study of
statistically stationary isotropic turbulence. Consistency with homogeneous shear flow, or with any problem in which  the forcing length scale increases as a power law or exponential in time \cite{Bob&Tim}, requires $C_{\epsilon 1} < C_{\epsilon 2}$.

Choosing $n\geq 2$ should 
improve the predictions by introducing the possibility of spectral imbalance, a necessary requirement if the same model is to be applied to both forced and decaying turbulence 
\cite{Bos2007-2}; imbalance is possible because
the partial fluxes $f_i$ with $1\leq i\leq n-1$ and the dissipation 
$\epsilon=f_n$ are independent.

The unsteady behavior of this model is now assessed as follows. 
Beginning with a steady state with shell energies $\bar{e}_i$ such that 
$\bar{p}=\bar{f}_i=\bar{\epsilon}$, the periodic perturbation
$\tilde{p}\cos(\omega t)$ is added to the forcing. Describing the
periodic response in terms of
complex amplitudes $\hat{k}(\omega)$ and 
$\hat{\epsilon}(\omega)$,
so that $\tilde{k}(\omega)=|\hat{k}(\omega)|$ and
$\tan\phi_k(\omega) = \Im\hat{k}(\omega)/\Re\hat{k}(\omega)$
with the obvious analogs for $\epsilon$,
the equations for the periodic part of the partial kinetic energies and local fluxes become
\begin{equation}\label{fnhat}
 i\omega{\hat{e}}_i=-(\hat f_i-\hat f_{i-1}), \qquad i\omega{\hat{f_i}}=-\frac{3}{2}\frac{\bar f_i}{\bar e_i}(\hat f_i-\hat f_{i-1}).
\end{equation}
The resulting linear system is easily solved analytically
for $\hat e_i(\omega)$ and $\hat f_i(\omega)$ in terms of the
periodic forcing perturbation $\tilde{p}$ and the parameters
$\bar e_i$ and $\bar f_i$. 

\begin{figure}
\setlength{\unitlength}{1\textwidth}
\includegraphics[width=0.45\unitlength]{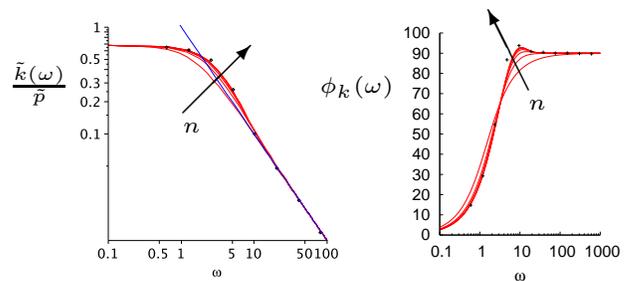}
\caption{Linear response functions $\tilde k(\omega)$ (left) and
$\phi_k(\omega)$ (right) for the multiple-scale model based on the 
Kovaznay spectral closure (\ref{Schiestel}). The 7 curves in each graph correspond to models with $n=1,2,..,7$ spectral shells and $n$ 
increases in the direction of the arrow. 
The theoretical amplitude prediction $\omega^{-1}$ is
indicated by a straight line. Also shown are the results of EDQNM simulations (symbols). The frequency  $\omega$ is normalized by the large scale frequency $\bar \epsilon/\bar k$.
\label{FigSchiestel}}
\end{figure}

Figure \ref{FigSchiestel} compares the results for the amplitude
$\tilde{k}(\omega)$ and phase shift
$\phi_k(\omega)$ for models with $n=1$ to $n=7$ wavenumber shells. Also shown are results obtained in EDQNM computations \cite{Bos2007-3} at Reynolds number 
$R_\lambda=1000$, with $R_\lambda\approx {15 R_L}^{1/2}$ and $R_L=(2\bar k/3)^{1/2}\bar L/\nu$. It has been shown \cite{Bos2007-3} that at low $\omega$, $\tilde k(\omega)$, should tend to a plateau and that
at large $\omega$, $\tilde k(\omega)$ follows a power-law
proportional to $\omega^{-1}$. 
Confirming the conclusion of Bos {\em et al.}\cite{Bos2007-3}, 
$\tilde{k}(\omega)$ is in reasonable agreement with EDQNM even for 
the two-equation model $n=1$, although this agreement improves significantly as the number
of wavenumber partitions $n$ increases.

The error in the two-equation model prediction of $\tilde{k}(\omega)$
is a too gradual transition from the plateau to
the $\omega^{-1}$ region. This defect appears more prominently in
the phase $\phi_k(\omega)$: 
although all models give the correct static and
frozen limits, the two-equation model transitions
much too gradually, and only the models with 
$n\geq 2$ are in close agreement with EDQNM.
The relatively rapid transition in the energy phase shift therefore
appears as a typical multiple-scale effect:
evidently, in this problem, the small scales are not
simply `slaved' to the large scales through a constant dissipation rate
as is assumed in a two-equation model; instead, they
are dynamically independent and have a strong effect on what is apparently
a purely large-scale property. 

\begin{figure}
\setlength{\unitlength}{1\textwidth}
\includegraphics[width=0.45\unitlength]{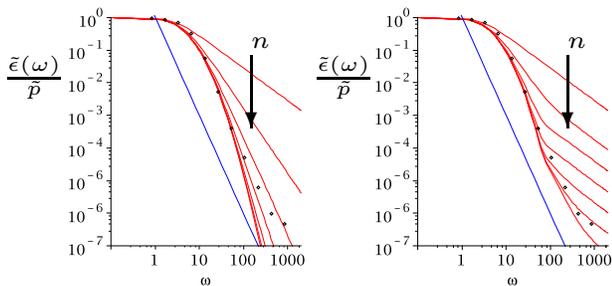}
\caption{
Frequency response function $\tilde \epsilon(\omega)$ for the model based on the Kovaznay closure
(\ref{fnhat}) (left) and for the model based on the Ellison closure (\ref{Ellison-f}) (right).
The 7 curves in each graph correspond to $n=1,2,..,7$ and $n$ increases in the direction of the arrow. 
The theoretical prediction $\omega^{-3}$ is indicated by the straight lines. 
Also shown are the results of EDQNM simulations (symbols).\label{FigEllison}}
\end{figure}

It has also been demonstrated\cite{Bos2007-3} that
the response function $\tilde \epsilon(\omega)$ follows at high $\omega$ a 
power-law proportional to $\omega^{-3}$ up to the Kolmogorov frequency 
$\omega_\eta\sim\bar \epsilon/\nu$.
For $\omega>\omega_\eta$, $\tilde \epsilon(\omega)$ becomes proportional to 
$\omega^{-1}$.  These results are shown for comparison in 
the graph on the left side of
Figure \ref{FigEllison} (the phase shift $\phi_{\epsilon}$ proves difficult to compute with any confidence because of the
extremely small amplitudes involved, therefore comparisons are omitted). 

The agreement of  $\tilde \epsilon(\omega)$ given by 
the multiple-scale model Eq. (\ref{fnhat})
with the EDQNM results is very good down to values 
$\tilde \epsilon /\tilde p=10^{-3}$ for $n>3$ (note that $\tilde k(\omega)$ and 
$\tilde \epsilon (\omega)$ are proportional to $\tilde p$, and $\tilde p$ is chosen 
unity without loss of generality). 
However for smaller values of $\tilde \epsilon /\tilde p$, the discrete model 
starts to diverge from the EDQNM results, especially for large $n$. 
It is very easily shown that for this model,
the leading order contributions at high $\omega$ are proportional to $\omega^{-n}$:
indeed, recursive solution of the equations for $\hat{f}_i$ in (\ref{fnhat})
show that
$ \tilde{f}_i \sim \omega^{-i}\tilde{p}.$
Thus, only if $n=3$ can we obtain $\tilde \epsilon(\omega)\sim \omega^{-3}$, but
this is the result of a coincidence, which disappears if the number of 
partition wavenumbers is increased. 
This difficulty reflects a limitation of the multiple-scale model: Eq. (\ref{dotF1}) implies
a linear first-order partial differential equation for $\hat{F}$ in which
disturbances in $F$ propagate along characteristics; this property is
probably significantly compromised by a 
finite dimensional approximate model.

It has been shown\cite{Bos2007-3} that nonlocal interactions are responsible for 
the $\omega^{-3}$ range. Nonlocal interactions are represented in Figure \ref{Cartoon} 
by dashed lines. 
Such interactions do not occur in the model Eqs. (\ref{Schiestel}) and (\ref{Schiestel-energy}),
because quantities in shell $i$ depend only on its nearest neighbor, shell $i-1$.
The absence of nonlocal interactions
will be even more significant in problems in which the role of nonlocality 
is greater, as in the Batchelor regime of the passive scalar 
\cite{Batchelor1959} and in MHD \cite{Pouquet1976}.

To address this problem, a new multiple-scale 
model will now be derived including the effect of nonlocal interactions. 
We start from a simple spectral model containing nonlocal interactions, 
due to Ellison \cite{Monin},
\begin{equation}
F(\kappa,t)=C\left(\int^\kappa_0\kappa^2E(\kappa,t)d\kappa \right)^{1/2}\kappa E(\kappa,t).
\end{equation}
Applying the same procedure as to the Kovaznay model, 
the partial energy balance Eq. (\ref{Schiestel-energy}) is unchanged, but
Eq. (\ref{Schiestel}) is replaced by
\begin{equation}
\dot f_i=-\frac{f_i}{e_i}(f_i-f_{i-1})-\frac{f_i}{2}\frac
{\sum_{p=1}^i{ \kappa_p^2(f_p-f_{p-1})}}
{\sum_{p=1}^i{\kappa_p^2 e_p}}.
\end{equation}
This expression contains the wavenumbers $\kappa_i$ because
nonlocal interactions, 
which depend on the spacing between the wave-number partitions,
have been retained. The specification of the $\kappa_i$ becomes 
part of the model: we will
use a logarithmic discretization 
with 
$\kappa_i=\kappa_1 r^{i-1}$ in which 
$r$ is a model parameter which determines the logarithmic gridsize. Using this
discretization, the ratio $\kappa_n/\kappa_1=r^{n-1}$, so that a large range of scales can be considered by increasing $r$. Note that if a linear discretization is used, $\kappa_n/\kappa_1= (n\Delta \kappa/1\Delta \kappa)= n$ so that the number of partitions for high Reynolds numbers becomes prohibitively
large. Using the logarithmic discretization, the
model for periodic forcing becomes
\begin{equation}
\dot{\hat{f_i}}=-\frac{\bar f_i}{\bar e_i}(\hat f_i- \hat f_{i-1})-\frac{\bar f_i}{2}\frac
{\sum_{p=1}^i{ r^{2(p-1)}(\hat f_p-\hat f_{p-1})}}
{\sum_{p=1}^i{ r^{2(p-1)} \bar e_p}},
\label{Ellison-f}
\end{equation}
with the same partial energy equations as in Eq. (\ref{fnhat}).
Again, when $n=1$ the model reduces to a two-equation model. 
For $n>1$ the model differs from the previous model through the interaction term which 
couples wave-number shell $i$ with all wavenumber shells $p=1,\cdots, i$.
The response functions depend on both $n$ and $r$, which will be chosen 
through a compromise 
between computational cost and precision. 

The results for $\tilde{\epsilon}(\omega)$ obtained from Eq. (\ref{Ellison-f}) with $r=3$ 
and $1\leq n\leq 7$ wavenumber partitions 
are shown in the graph on the right side of Figure \ref{FigEllison}. 
In the same figure, EDQNM results \cite{Bos2007-3} at
$R_{\lambda}=1000$ are shown. 
The agreement 
with EDQNM is very good down to values 
$\tilde \epsilon /\tilde p=10^{-3}$ for $3<n<7$, and for $n=7$, agreement is good down 
to $\tilde \epsilon /\tilde p =10^{-6}$.
We conclude that the model including nonlocal interactions 
Eq. (\ref{Ellison-f}) makes better predictions of $\tilde{\epsilon}$
than the model Eq. (\ref{fnhat}), in which nonlocal interactions are absent.
The predictions of Eq. (\ref{Ellison-f}) (not shown) for $\tilde k(\omega)$ and
$\phi_k(\omega)$ very nearly coincide
with the results obtained using Eq. (\ref{fnhat}).

To conclude, we have found that in the problem of periodically forced turbulence, 
a two-equation model only gives satisfactory predictions for 
$\tilde{k}(\omega)$ and $\phi_k(\omega)$ at asymptotically high and low frequencies. 
The predictions at intermediate frequencies
are improved by using a multiple-scale model based on the heuristic picture of stepwise
energy cascade, the Kovaznay model. In particular, this model  correctly predicts the rapid jump of the phase shift
$\phi_k(\omega)$ between the static and frozen limits.
The multiple-scale model
based on the Ellison closure includes nonlocal effects, and leads 
to better agreement with EDQNM, including
the high Reynolds number $\omega^{-3}$ scaling range\cite{Bos2007-3}.
Both models give practically indistinguishable results for the amplitude and 
phase of the oscillating kinetic energy, which
is not strongly influenced by nonlocal interactions. 
Our approach suggests how one might construct
reduced order models for phenomena dominated by significant
nonlocal interactions, like the Batchelor range of the passive scalar and
some cases of MHD.

We would like to acknowledge the interesting and thoughtful comments of
the referees, which led to significant modifications of the paper.

\end{document}